\documentstyle[11pt,newpasp,twoside,epsf]{article}
\def    \bOmega	{\vec \Omega}
\def 	  \be 		{\begin{equation}}
\def	  \ee		{\end{equation}}

\def  \cm  {\rm cm}

\def\edcomment#1{\iffalse\marginpar{\raggedright\sl#1\/}\else\relax\fi}
\reversemarginpar
\begin{document}

\title{Translational
Velocities and Rotational Rates of Interstellar Dust Grains}

\author{A. Lazarian \& Huirong Yan}

\affil{Department of Astronomy, University of Wisconsin, 475 N. Charter
St., Madison, WI 53706}

\begin{abstract}

Interstellar dust grains exhibit complex dynamics which is essential for
understanding many key interstellar processes that involve
dust, including grain alignment, grain growth, grain shattering
etc. Grain rotational and translational
motions are affected not only by gaseous collisions, but also by 
interactions with ions, photons, magnetic fields etc. Some of those
interactions, e.g. interactions of ions with the dipole electric moment
of dust grains,  require the quantum nature of the process to be accounted
for. Similarly,
coupling of rotational and vibrational degrees of freedom in a grain
happens due to  relaxation processes, among which the process
related to nuclear spins frequently is the dominant one. 
This coupling
modifies substantially both the dynamics of rotational and translational
motions by inducing grain flips. The flipping averages our certain
systematic torques and forces that act on grains. As the rate of
flipping is larger for smaller grains, these grains can be ``thermally
trapped'', i.e. rotate at a 
thermal rate in spite of the presence of systematic torques. Moreover,
a subset of small grains with high dipole moments may rotate at subthermal
rates due to high damping arising from grain emission in microwave range
of frequencies. Translational and rotational dynamics of grains
is interrelated. For instance, both rotation and gas grain relative 
speeds affect grain alignment as well as 
determine grain size distribution and structure.
Translational dynamics of grains is mostly dominated by
grain interactions with magnetohydrodynamic turbulence. Efficient
turbulent mixing of dust grains limits the degree to which grains
of different sizes may be segregated in space.  

\end{abstract}

\section{Why do we care?}

	Dust is an important constituent of the interstellar medium (ISM).
It interferes with observations in the optical range, but provides
an insight to star-formation activity through far-infrared radiation.
As dust gets aligned in the external magnetic field (Hall 1949;
Hiltner 1949) it traces the magnetic field via emission and 
extinction polarization (see reviews by Hildebrand et al. 2000;
Lazarian 2003).  
Grain alignment is a property of rotating grains. Fast
rotation makes grains less susceptible to disorientation by collisions with gas atoms,
but it also limits the efficiency of some alignment mechanisms.
In addition, fast rotating grains are the sources of microwave
emission that interferes with measurements of Cosmic
Microwave Background (CMB) intensity and polarization (
Draine \& Lazarian 1998a; see also recent
review by Lazarian \& Finkbeiner 2003).

The basic properties
of dust (optical extinction, dust chemistry, 
heating of the ISM etc) 
strongly depend on its size distribution (Biermann \& Harwit 1980; 
O'donnell \& Mathis 1997).
The latter evolves As a result of grain collisions, whose frequency
and consequences depend on grain relative velocities (Draine 1985).

Rotational and translational motions are not independent. For instance,
the difference in the number of active sites of H$_2$ formation
over the parts of the grain
perpendicular to the axis of grain rotation can result in uncompensated
force accelerating the grain (Purcell 1979, Lazarian \& Yan 2002, henceforth LY02). However, 
if a grain is undergoing frequent flips due to thermal fluctuations
(Lazarian \& Draine 1999a, henceforth LD99a and \S 2) the thrust will change direction
and be averaged out.
On the other hand, grain alignment presents a case when translational
motion affects grain rotation. For instance, mechanical alignment of grains
(see Lazarian 2003 and references therein) induces
higher rotational rates if grains rotate thermally initially.

Damping of rotational and translational motions have many similarities.
For instance, the interaction of grains having dipole moments
with passing ions is important for damping of both types of motions.
For the case of damping of translational motion the treatment of
the process is given in Yan, Lazarian \& Draine (2003, henceforth YLD03). 
Quantum
cut-off discussed in \S 4 is applicable to this case as well.  

In what follows we discuss how dust grains rotate (\S 2),
at what rate they rotate (\S 3), and whether quantum effects
are important for grain rotation (\S 4). We describe the
translational motion of grains in \S 5. In \S 6 we discuss
astrophysical implications of grain dynamics and provide
the summary in \S 7.

\section{Do grains rotate about their axis of major inertia?}

Originally this question was asked in relation with the theory
of grain alignment (see  Fig.~1a and discussion in Lazarian 2003). 
To produce the observed starlight polarization, grains must be aligned
with their long axes perpendicular to magnetic field. This
involves alignment not only of grain angular momentum ${\bf J}$ in
respect to the external magnetic field ${\bf B}$, but also the alignment
of grain long axes in respect to ${\bf J}$. Jones
\& Spitzer (1967) assumed a Maxwellian distribution of angular
momentum which favored the preferential alignment of ${\bf J}$
with the axis of the maximal moment of inertia (henceforth
axis of major inertia). Purcell (1979, henceforth P79) later considered 
grains rotating much faster than the thermal velocities (see \S3)
and showed internal
dissipation of energy in a grain will make grains rotate about
the axis of major inertia.

Arguments in P79 can be easily understood. Indeed, for an oblate 
grain (see Fig.~1b) with angular momentum $J$ the energy can be
written as
\begin{equation}
E(\theta)=\frac{J^2}{I_{max}}\left(1+\sin^2\beta(h-1)\right)~~~,
\label{1}
\end{equation}
where $h=I_{max}/I_{\bot}$ is the ratio of the maximal to minimal
moments of grain inertia. Internal forces cannot change the angular
momentum, but it is evident from eq.(1) that the energy can be
decreased by aligning the axis of maximal inertia along ${\bf J}$,
i.e. by decreasing $\beta$.  
P79 discussed two possible causes of internal dissipation,
the first one related to the well known inelastic relaxation (see also Lazarian \& Efroimsky 1999), the second is
due to the mechanism that he discovered and termed ``Barnett relaxation''.

We remind the reader that the
Barnett effect is converse of the Einstein-de Haas effect.
If in Einstein-de Haas effect a paramagnetic body starts rotating
 during remagnetizations as its flipping 
electrons transfer the angular momentum (associated with their spins)
 to the
lattice, in the Barnett effect
the rotating body shares its angular momentum with the electron
subsystem  causing magnetization. The magnetization
is directed along the grain angular velocity and the value
of the Barnett-induced magnetic moment is $\mu\approx 10^{-19}\omega_{(5)}$~erg
gauss$^{-1}$ (where $\omega_{(5)}\equiv \omega({\rm s}^{-1})/10^5$)\footnote{Therefore
the Larmor precession has a period $\tau_L\approx 3\times 10^6 B_{(5)}^{-1}$~s (where $B_{5}^{-1}$) and 
the magnetic field defines the axis of alignment (see more details in
Lazarian 2003)}.

\begin{figure} [h!t]
{\centering \leavevmode
\epsfxsize=1.9in\epsfbox{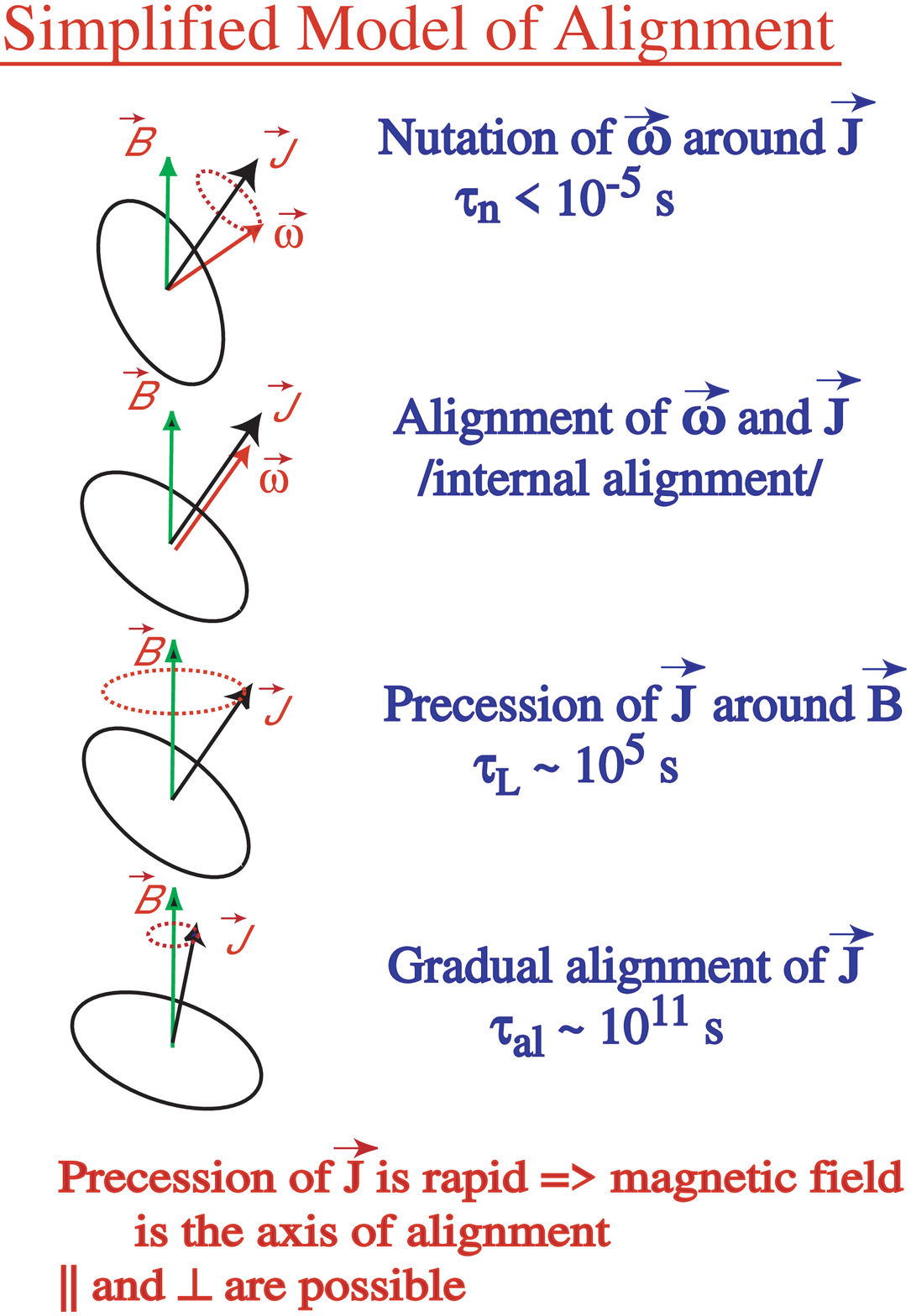}
\hfil
 \epsfxsize=2.1in\epsfbox{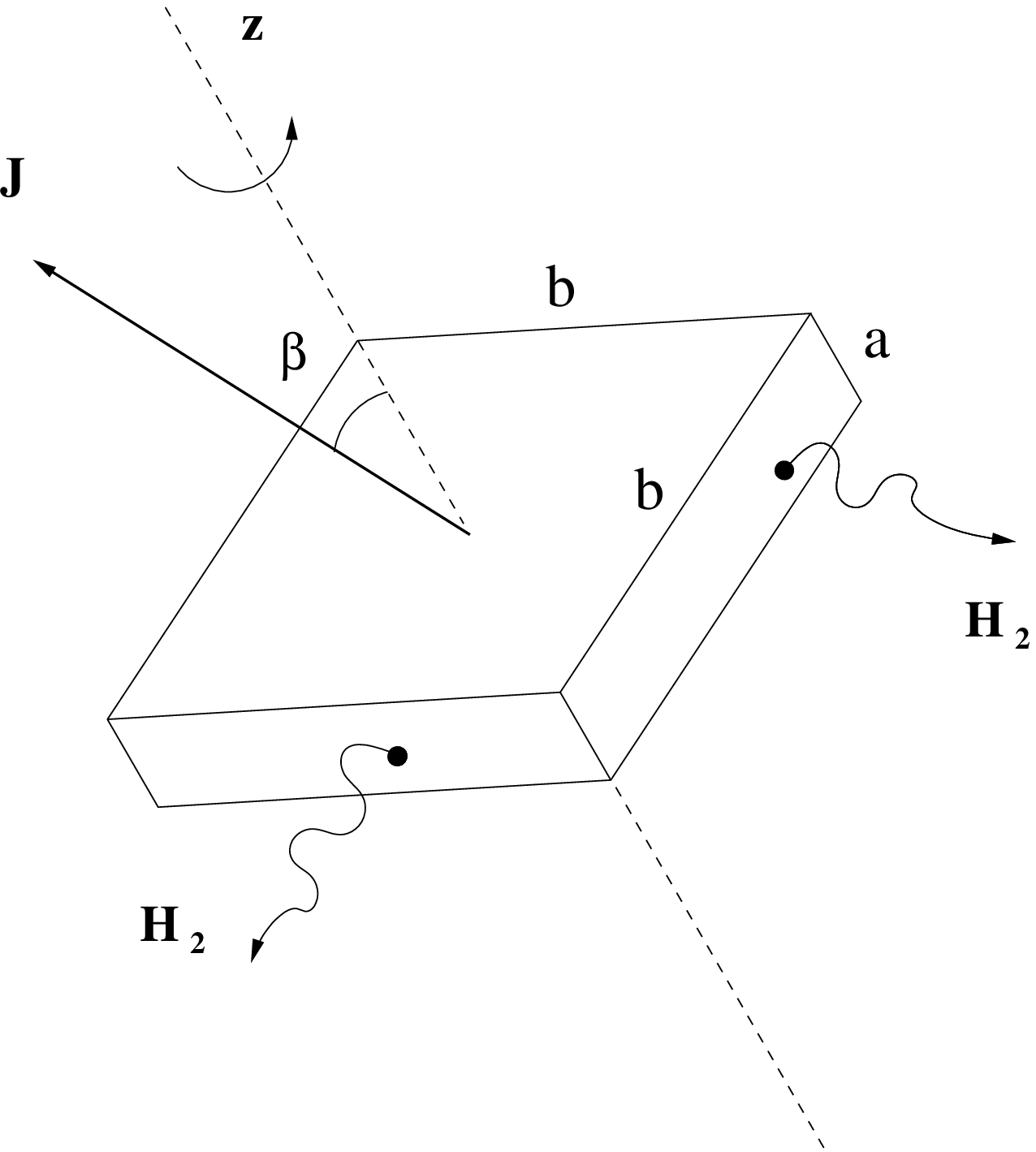} }
\caption{{\it Left panel}-- Grain alignment implies several alignment
processes acting simultaneously and spanning many time scales
(shown for $10^{-5}$~cm grain in cold interstellar gas). The
rotational dynamics of a grain is rather complex. The internal alignment
introduced by Purcell (1979) was thought to be slower than precession
until Lazarian \& Draine (1999b, henceforth LD99b) showed that it happens $10^6$ times
faster when relaxation through induced by nuclear spins is accounted for
(approximately $10^4$~s for the $10^{-5}$~cm grains). {\it Right panel}--
Grain rotation arising from systematic torques arising from H$_2$ formation
(P79). In the presence of efficient internal relaxation the angle $\beta$
between the axis of maximal moment of inertia and $\bf J$ is small is
grain is rotating at suprathermal rates ($E_{kinetic}\gg kT_{grain}$).}
\end{figure}

The Barnett relaxation process may be easily understood. We know that a 
freely rotating grain preserves the direction of
${\bf J}$, while angular velocity precesses about 
${\bf J}$ and in grain body axes. The 
``Barnett equivalent magnetic field'', i.e. the equivalent external
magnetic field that would cause the same magnetization of the grain  
material, is $H_{BE}=5.6 \times10^{-3} \omega_{(5)}$~G. Due to
the precession of angular velocity the ``Barnett equivalent magnetic field''
changes in grain axes. This causes remagnetization accompanied
by the inevitable dissipation. As a result $E(\theta)$ and therefore
$\theta$ decreases.  

\begin{figure}
{\centering \leavevmode
\epsfxsize=2.3in\epsfbox{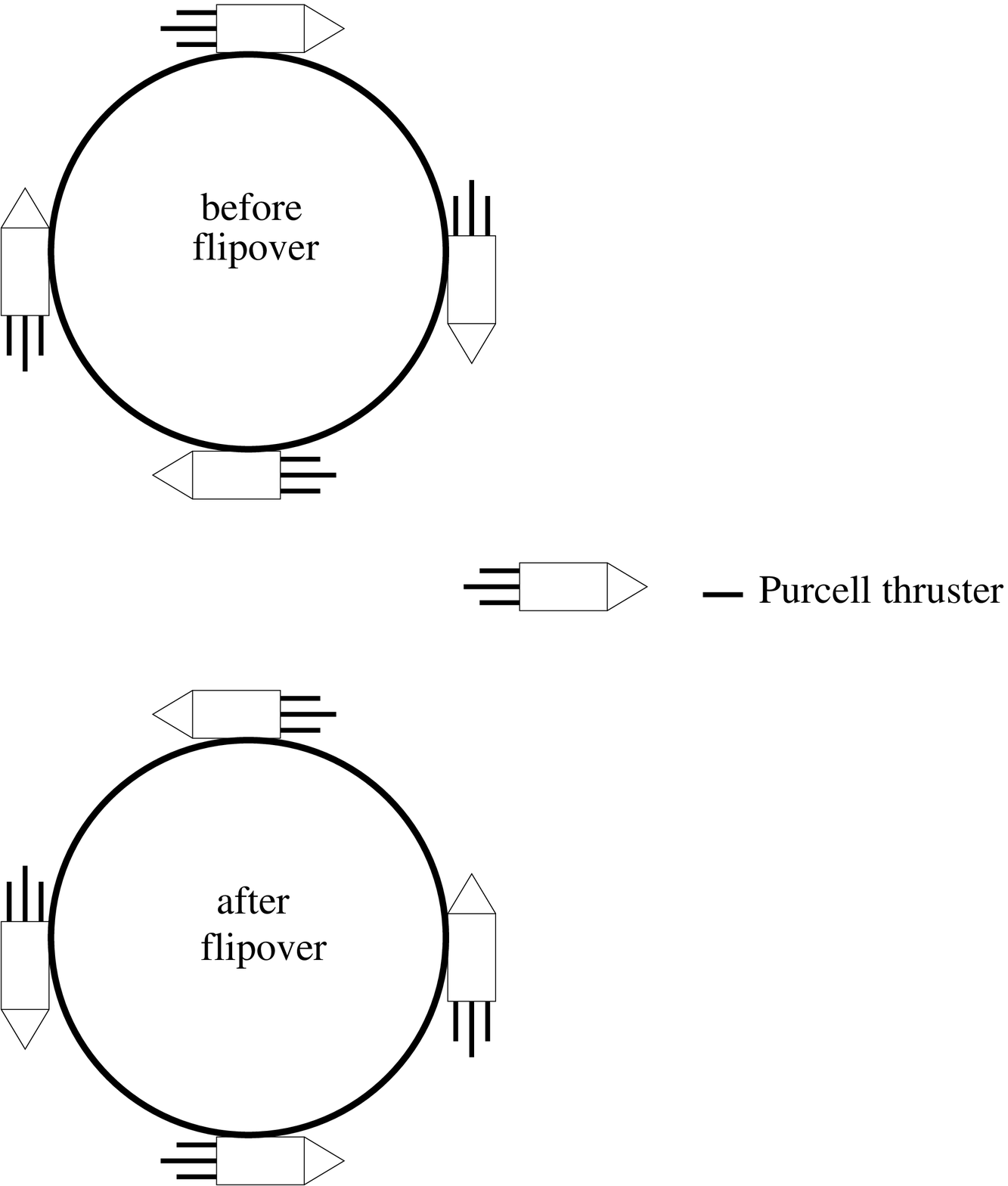}
\hfil
 \epsfxsize=2.0in\epsfbox{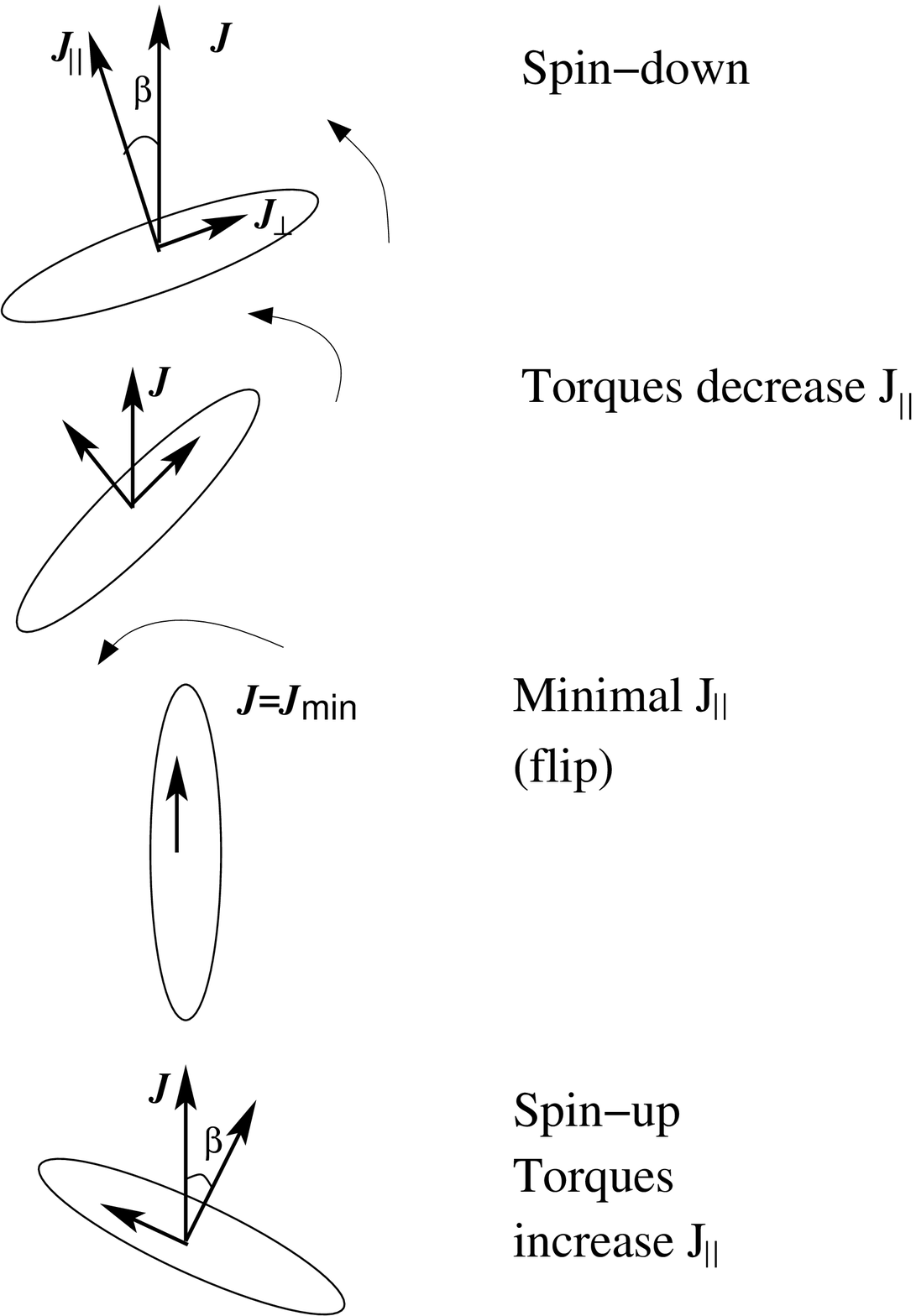} }
\caption{{\it Left panel}-- Schematic of a grain affected by Purcell's
 torques before and after a flipover event. As the grain flips
the direction of torques changes. H$_2$ formation sites act as thrusters
in the figure. {\it Right panel}-- Regular crossover event as it described
by Spitzer \& McGlynn (1979). The systematic torques decrease the
amplitude of ${\bf J}$ component 
parallel to the axis of maximal inertia to zero while preserving the other
component, $J_{\bot}$. If $J_{\bot}$ is small then 
the grain is susceptible to randomization during crossovers. The direction
of ${\bf J}$ is preserved in the absence of random bombardment.}
\end{figure}

The Barnett relaxation happens on the scale 
$t_{Bar}\approx
4\times 10^7 \omega_{(5)}^{-2}$~sec, very short compared to 
the time $t_{gas}$ over which randomization through gas-grain
collisions takes place. As a result, models of interstellar polarization
developed since 1979 have often assumed that Barnett dissipation
aligns $\bf J~ {\it perfectly}\/$ with the major axis of inertia.
However, Lazarian (1994, henceforth L94) has pointed out that this is a
poor approximation if the grains rotate with thermal kinetic energies:
thermal fluctuations in the Barnett magnetization will excite
rotation about all 3 of the body axes, preventing perfect
alignment unless the rotation is suprathermal or the grain solid
temperature is zero. The exact analysis of the problem was
given in Lazarian \& Roberge (1997),  
where the distribution of $\theta$ for a freely
rotating grain was defined through the Boltzmann distribution
 $\exp(-E(\theta)/kT_{grain})$, where $T_{grain}$ is the grain
temperature.

\section{How fast do the grains rotate?}

Earlier work in the field assumed Brownian grain rotation
with the effective temperature equal to the mean of
the grain and gas temperatures (see Jones \& Spitzer 1967).
The complexity of grain rotation was realized only later.
Purcell (1975; 1979) realized that grains may rotate at a much faster
rate resulting from systematic torques.  P79
 identified three separate systematic torque
mechanisms: inelastic scattering of impinging atoms when gas and grain
temperatures differ, photoelectric emission, and H$_2$ formation on
grain surfaces (see Fig.~1b); we will refer to these below as "Purcell's torques".
The latter was shown to dominate the other two for 
typical conditions
in the diffuse ISM (P79).  The existence of systematic H$_2$ torques
is expected due to the random distribution over the grain surface of
catalytic sites of H$_2$ formation, since each active site acts as a
minute thruster emitting newly-formed H$_2$ molecules.

Independent of Purcell, Dolginov (1972) and Dolginov
\& Mytrophanov (1975) identified radiative torques as the
way of spinning up a subset of interstellar grains.
``Helical'' grains would scatter differently left and right
polarized light and therefore ordinary unpolarized light would
spun them up.
        This subset of ``helical'' grains was believed
	to be somewhat limited to special shapes/materials.
	This work did not make much impact to
	the field until
        Draine \& Weingartner
        (1996) showed that grains of arbitrary irregular shapes get
	spun up when their size is of the order of the starlight
	wavelength.

The arguments of P79 in favor of suprathermal rotation were so clear
and compelling that other researchers were immediately convinced that
interstellar grains in diffuse clouds should rotate suprathermally.
Purcell's discovery was of immense importance for grain alignment.
Suprathermally rotating grains remain subject to gradual alignment by
paramagnetic dissipation (Davis \& Greenstein 1951), but due to their
large angular momentum are essentially immune to disalignment by
collisions with gas atoms.

Spitzer \& McGlynn (1979, henceforth SM79) showed that suprathermally
rotating grains should be susceptible to disalignment only during
short intervals of slow rotation that they called
``crossovers'' (see Fig~2, right). During a crossover the grain slows down, flips
and is accelerated again. Crossovers 
	are due to various grain surface processes that change the 
	direction (in body-coordinates) of the systematic torques.
Therefore for sufficiently infrequent crossovers suprathermally
rotating grains will be well aligned with the degree of alignment
determined by the time between crossovers, the degree of correlation
of the direction of grain angular momentum before and after a
crossover (SM79), and environmental conditions (e.g., magnetic field
strength $B$).

\begin{figure}
{\centering \leavevmode
\epsfxsize=2.2in\epsfbox{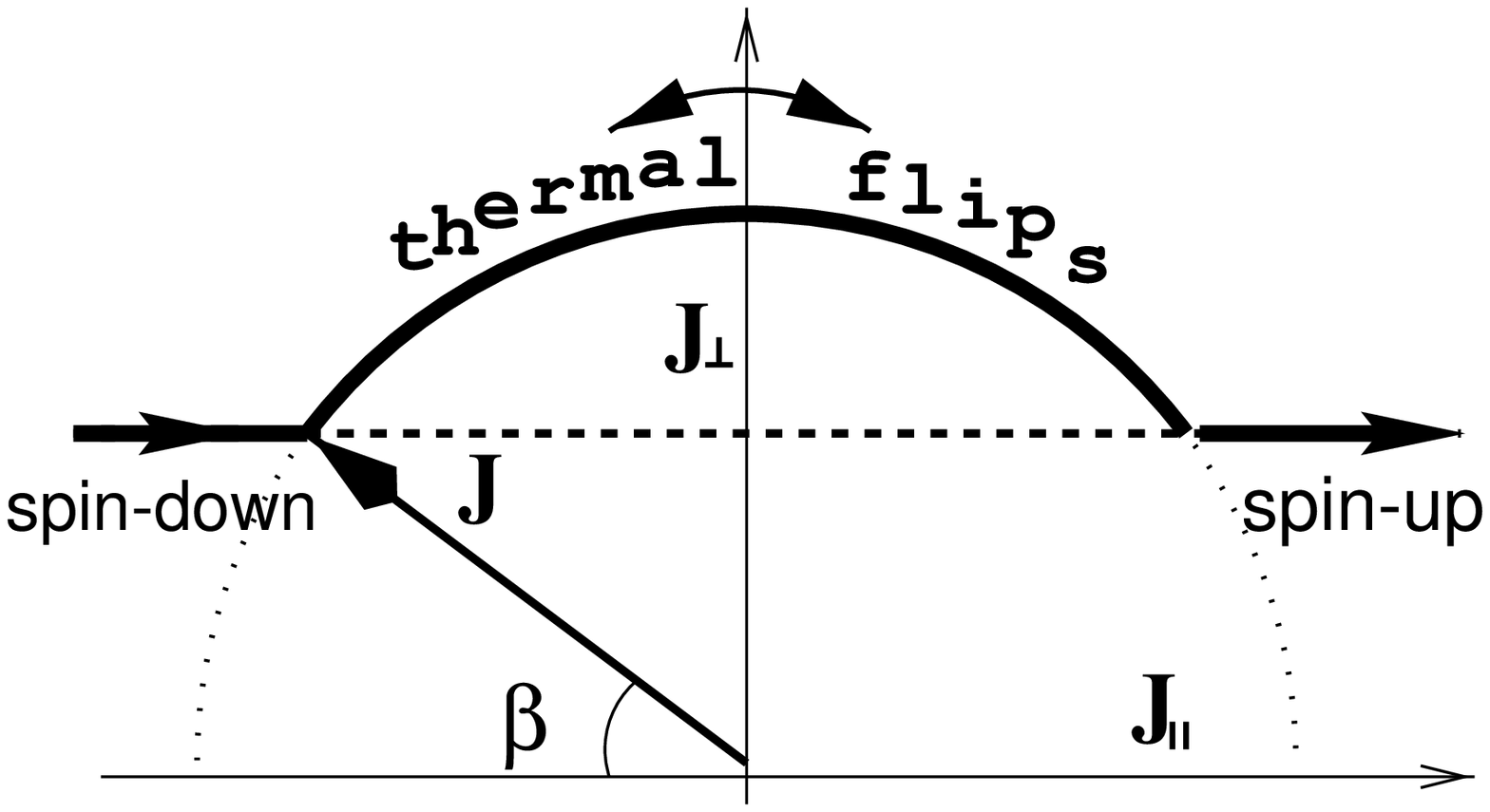}
\hfil
\epsfxsize=1.9in\epsfbox{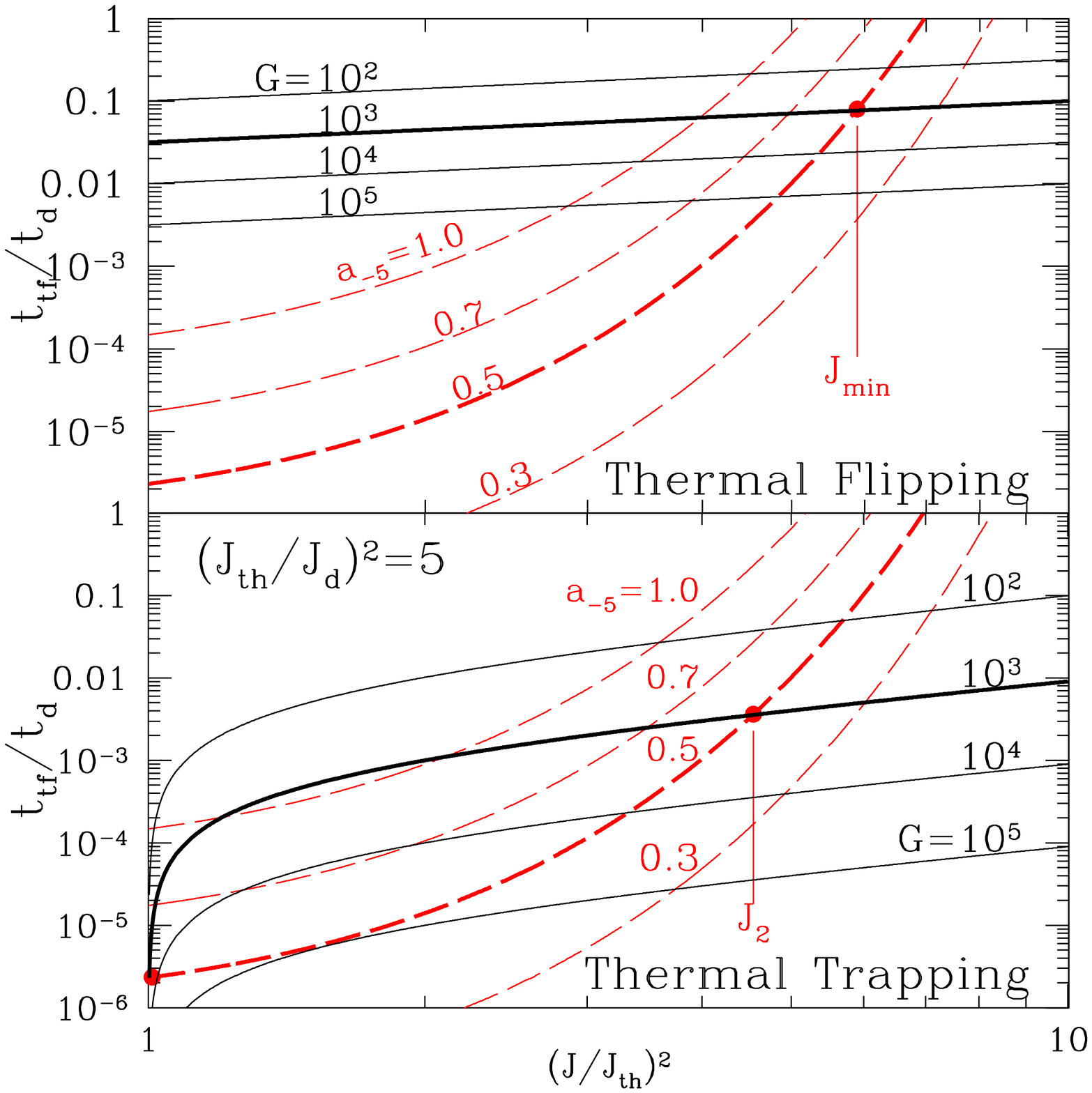}}
\caption{{\it Left panel}--
Grain trajectory on the $J_\perp$ -- $J_\parallel$ plane, where
        $J_\perp$ and $J_\parallel$ are components of $\bf J$ 
        perpendicular or parallel
        to the grain's principal axis of largest moment of inertia.
        The solid trajectory shows a ``thermal flip'', while the broken line
        shows the ``regular'' crossover which would occur in the absence of
        a thermal flip. From LD99a. {\it Right panel}-- 	
       Top: Thermal flipping to damping ratio as a function of
       $J/J_{thermal}$ for grains of given size
	[broken lines, labeled by $a_{-5}\equiv a(cm)/10^{-5}$] and
	for grains with a given value of systematic torques 
        [solid lines, labeled by $G$].
	Dot shows $J_{\rm min}= \dot{J}\cdot t_{tf}$ 
        for flipping-assisted crossover of $a_{-5}=0.5$
	grain with $G=10^3$.
	Bottom: Thermal trapping for grains of given size [broken lines,
	labeled by $a_{-5}$], and given value of torques 
	[solid lines, labeled by $G$].  From LD99a.}
\end{figure}

The original calculations of SM79 obtained only marginal correlation
of angular momentum before and after a crossover, but their analysis
disregarded thermal fluctuations within the grain material.  
Indeed, if the alignment of ${\bf J}$ with the axis of major inertia
is perfect all the time through the crossover
 the absolute value of $|{\bf J}|$ passes through zero during
the crossover. Therefore gas collisions and recoils from
nascent $H_2$ molecules would {\it completely} randomize
the initial and final directions of ${\bf J}$ during the crossover. 
Thermal fluctuations partially decouple ${\bf J}$ and the axis of
major inertia (see \S2). As a result the minimal
value of $|{\bf J}|$ during a
crossover is equal to the component of ${\bf J}$ perpendicular
to the axis of major inertia. This value is approximately 
$J_{thermal}\sim (2kT_{grain} I_{max})^{1/2}$ and the
randomization during
a crossover decreases (Lazarian
\& Draine 1997, henceforth LD97). LD97 obtained 
a high degree of correlation of angular momentum direction
before and after the crossover for
grains larger than a critical radius $a_c \approx 1.5\times10^{-5}$cm, 
the radius for which the
time for internal dissipation of rotational kinetic energy is equal to
the duration of a crossover.  

What would happen for grains that are smaller than $a_c$? Lazarian \&
Draine (1999a, henceforth LD99a) showed that instead of following the
phase space trajectory prescribed by SM79 theory along which
$J_{\bot}$ is approximately constant, while the component of ${\bf J}$
 parallel to the axis of maximal inertia $J_{\|}$ changes sign, 
the grains undergo flipovers (see Fig2a) during which $|J|$ does not change
(see Fig3a). If these flipovers repeat; grains get
``thermally trapped'' (LD99a and Fig3b). This process can be understood in the 
following way. For sufficiently small $|J|$ the rate of flipping
$t_{tf}^{-1}$ becomes large. Purcell's torques change sign as
grain flips. As a result, grain $J^2$ undergoes a random walk over
the time of grain rotational damping $t_d$. If $J_{th}$ is the
momentum of a thermally rotating grain, the expected value
of the mean squared angular momentum $\langle J^2\rangle$
will be approximately $J^2_{th}+(G-1)J_{th}^2 t_{tf}/(t_d+t_{tf})$,
where $G$ is defined by the systematic Purcell's torque being $(G-1)^{1/2}J_{th}/t_d$.
As $t_{tf}(J^2)$ is a non-linear function of angular momentum the
solution for the angular momentum is bistable. For sufficiently
small initial $J^2$ less than the critical
value $J^2_{cr}$, it tends to $\approx J_{th}^2$, while 
for of $J^2>J^2_{cr}$ it tends to its suprathermal value
predicted by Purcell i.e. $\approx G J_{th}^2$. A suprathermally
rotating grain has chances to become thermally trapped and stay
in the trapped state for $\approx t_d \exp[(J_{cr}/J_{th})^2]$
till fluctuations in the value of the angular momentum rescue
the grain from the trap.
As a result a substantial part of grains smaller
than $a_{cr}$ will not rotate at high rates predicted by P79
even in spite of the presence of systematic torques that are fixed in body coordinates. 
Thermally rotating grains are subject to randomization by gaseous
collisions and are marginally aligned for typical interstellar fields
(see Lazarian 1997a; Roberge \& Lazarian 1999). 
Other consequences of thermal trapping and
the values of $a_{cr}$ that follow from processes different from
the Barnett relaxation are
discussed below.
The LD99a model gives plausible predictions, but a more comprehensive
quantitative study is necessary. We are aware of such a study being done
by Roberge \& Ford (in preparation; see also the contribution by Roberge (2004)
to this volume).

Thermal trapping limits the range of grain sizes which can be spun
up by Purcell's torques. One may expect that radiative torques are
not much influenced by thermal trapping as they are not fixed in
the grain axes and may not alter their direction due to grain flipping.
A quantitative study by Draine \& Weingartner (1996) shows that for
typical interstellar spectra grains with sizes larger than
$5\times 10^{-6}$~cm can be spun up by radiative torques\footnote{In
the vicinity of stars with UV excess smaller grains can be spun up
as well.}  This means that in
ISM grains larger than $5\times 10^{-6}$~cm
rotate suprathermally\footnote{An important point here is that radiative 
torques
may not be dominant in terms of their absolute value, but still
they can rescue grains from thermal trapping.}, while small grains
may rotate thermally.

Can grains rotate subthermally, i.e. with velocities which are substantially
smaller than those of Brownian rotation? The answer to this is positive.
Draine \& Lazarian (1998a, henceforth DL98) 
showed that grains with sizes less than
$10^{-7}$~cm may be efficient emitters of microwave radiation
provided that they have dipole moments. The emission from those grains
can account
for the so-called Foreground X (De Oliveira-Costa et al. 2002) 
observed in the range 10-100~GHz (see Fig.~4a). 
The back-reaction from radiation 
slows down grain rotation. As a result, detailed calculations in
DL98 showed that emitting grains may rotate
at substantially subthermal velocities, resulting in considerably less microwave emission than had been estimated by Ferrara \& Dettmar (1994), who assumed Brownian rotation.

\section{Are quantum effects important for grain dynamics?}

Dust grains are essentially macroscopic bodies. The rotational quantum
number $J/\hbar>100$ even for the smallest
grains (DL98). Therefore the initial reaction is that grain dynamics can
always be described using classical physics. This is not true, however.

DL98 showed that to describe the interaction of a grain with a dipole 
moment with ions one must account for the quantum nature of the interaction.
Indeed, the pioneering work by Anderson \& Watson (1993) assumed
 that ions within the Debye screening length 
interact with the grain dipole moment.
 
Calculations in DL98 showed that the more severe
limitations on which ions can interact with grain dipole moment
come in case of sufficiently small grains 
from the requirement that the angular momentum transfered
in an individual\footnote{Excitation of plasma waves as an additional
source of damping was considered in Ragot (2002). However, it is possible
to show that for small interstellar grains considered in DL98 collective
plasma effects are negligible.} interaction should be multiples of $\hbar$.   

Another case when quantum mechanics is important is related to
the Barnett effect.
The Barnett effect is a quantum effect, as it involves orientation
of electron spins, which are quantum objects. P79 noted that an
analog of the Barnett effect exists for nuclear spins. If a rotating body has initially an
equal number of nuclear spins directed
parallel and anti-parallel to the angular velocity $\bOmega$, 
it can decrease its kinetic energy, at constant total angular
momentum $\bf J$, if some of the angular momentum is transfered to the
nuclear spin system. 
Increasing the projection of the nuclear angular
momentum along $\bf J$ by $+\hbar$
(at constant $J$)
reduces the rotational kinetic energy by $\hbar \Omega$.
If the rotating body is allowed to come into thermal equilibrium (without
exchanging angular momentum) with
a heat reservoir of temperature $T_{\rm dust}$ then particles of spin $S$
develop a net alignment per particle
\be
\frac{\sum_{m=-S}^Sm\exp(m\hbar\Omega/kT_{\rm dust})}
	{\sum_{m=-S}^S\exp(m\hbar\Omega/kT_{\rm dust})} ~.
\label{eq:tanh}
\ee
Note that this
does not depend on the magnetic moment 
$\mu$. 

As the number of 
parallel and antiparallel spins becomes different the body develops 
magnetization. 
The relation between $\Omega$ and the 
strength of the ``Barnett-equivalent'' magnetic
field 
$H_{\rm BE}^{\rm (n)}$
(P79) 
that would cause the same 
nuclear
magnetization (in a nonrotating
body) is given by 
\be
{\bf H}_{\rm BE}^{(\rm n)}=\frac{\hbar}{g_{\rm n}\mu_{\rm N}}\bOmega~~~,
\label{H}
\ee
where $g_{\rm n}$ is the so-called nuclear $g$-factor (see Morrish 1980),
and $\mu_{\rm N}\equiv e\hbar/2m_{\rm p}c$ is the nuclear magneton,
smaller than the Bohr magneton by the electron to proton mass
ratio, $m_{\rm e}/m_{\rm p}$. A striking feature of eq.~(3)
is that the Barnett-equivalent magnetic field is inversely proportional
to the species magnetic moment. As grain tumbles
the magnetization changes in grain body coordinates
and this causes paramagnetic relaxation. This relaxation is proportional
to $\chi_N^{\prime\prime} (\omega) H^2_{BE}$ (where $\chi_N^{\prime\prime}$ is the imaginary part of the nuclear contribution to the susceptibility) and is approximately $10^6$ times
faster for nuclear moments than for their electron counterparts\footnote{
Obviously enough this sort of arguments fails when the rate of rotation
is larger than the rate at which nuclear spins interact. The latter
depends on the concentration of nuclear spins. For the ISM dust
LD99b concluded that the nuclear relaxation is dominant for thermally
rotating grains larger than $\sim 5\times 10^{-6}$~cm.}.

\begin{figure}
{\centering \leavevmode
\epsfxsize=2.3in\epsfbox{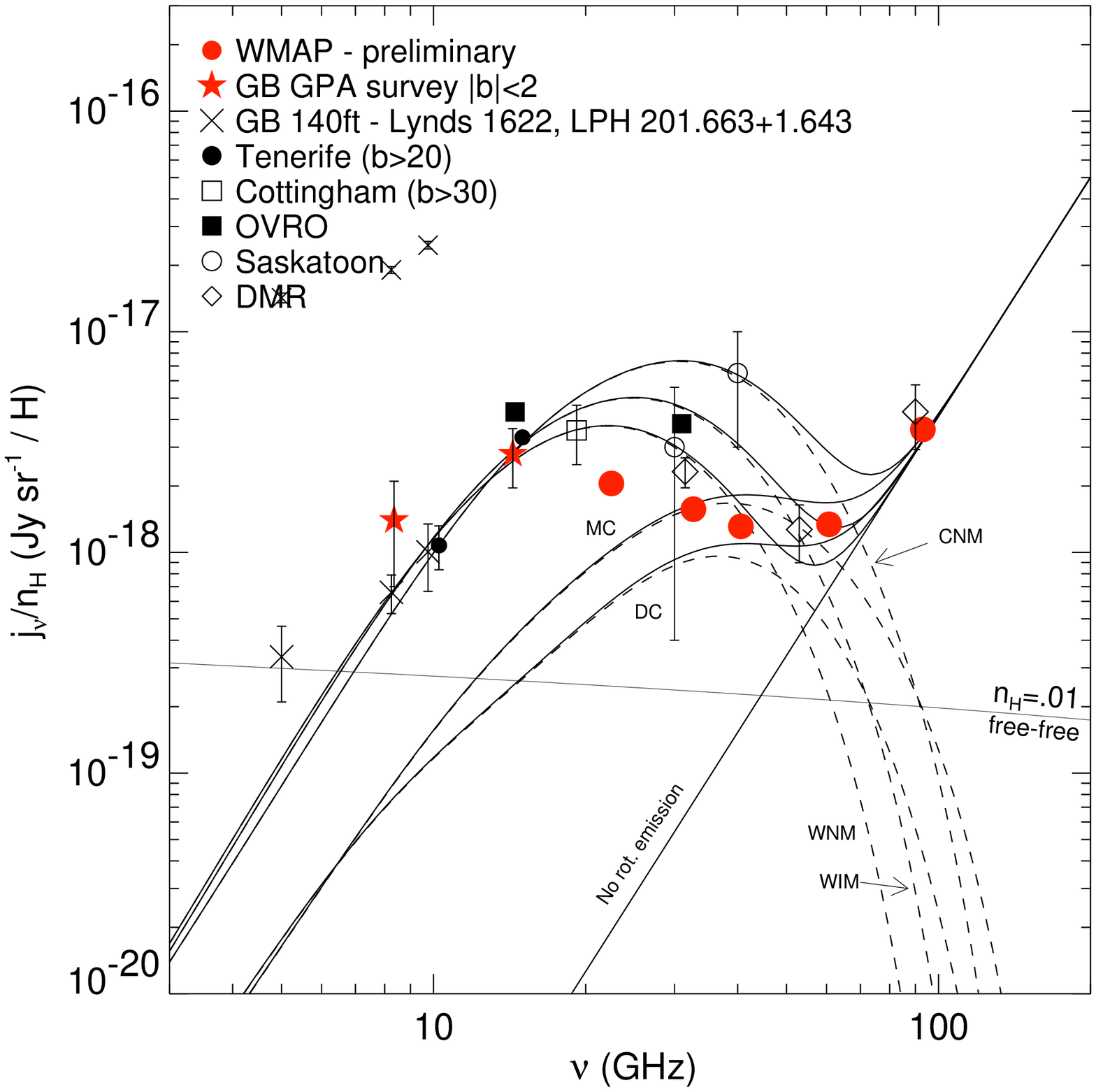}
\hfil
\epsfxsize=2.1in\epsfbox{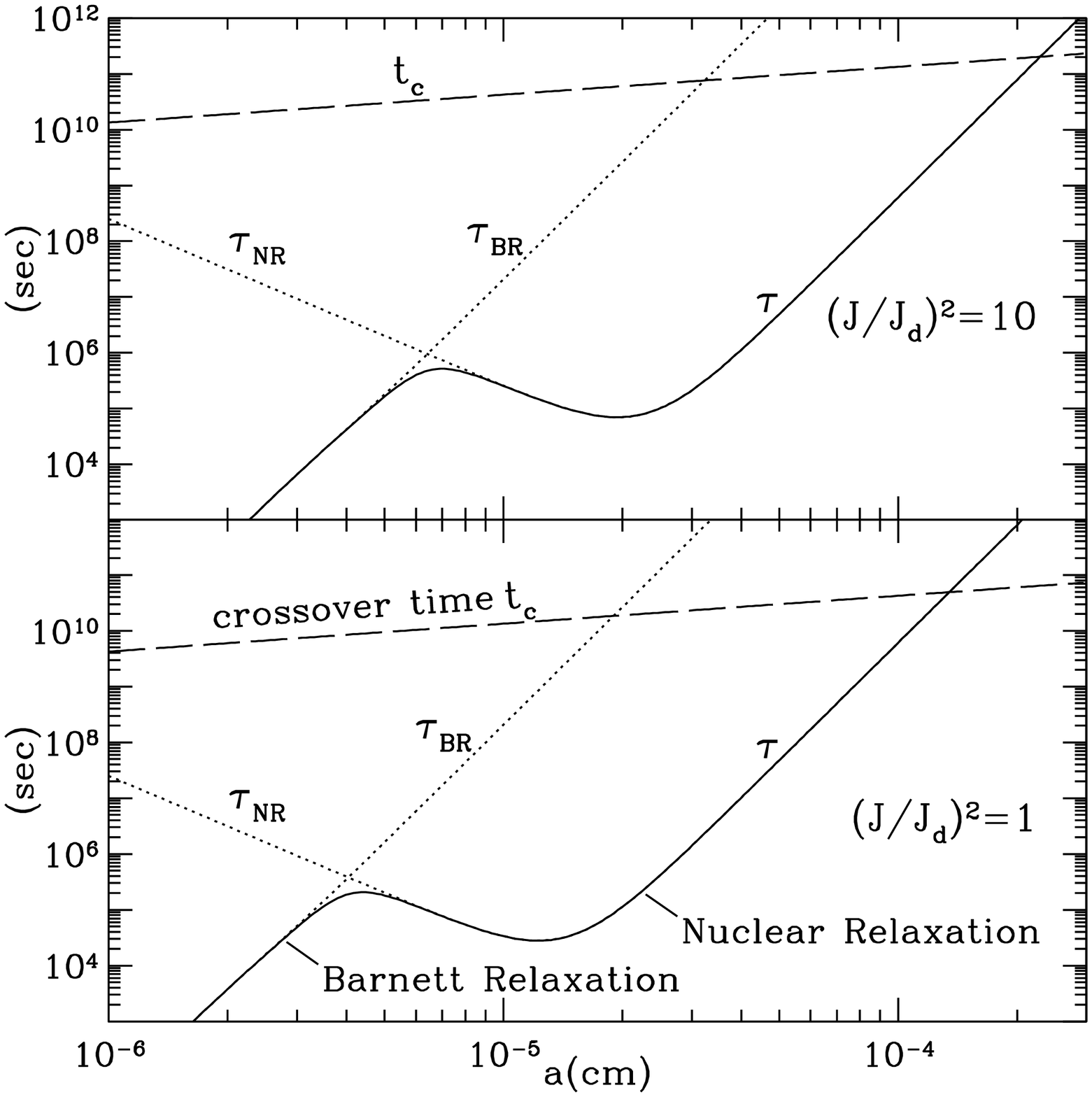}}
\caption{{\it Left panel}--
	Emission of spinning grains for different phases of interstellar
	media in the model by Draine \& Lazarian (1998b) against observational
	data (WMAP data is by D. Finkbeiner). From Lazarian \& Finkbeiner 2003.
	{\it Right panel} -- Time $\tau$ for alignment of $\hat{a}$ with 
	$\bf J$
	for a freely rotating
	grain, due to dissipation associated with nuclear spins
	(nuclear relaxation) and electron spins
	(Barnett effect).
	Results are shown for grains with $(J/J_d)^2=1$ (bottom)
	and 10 (top).
	Also shown is the ``crossover time'' $t_c = J/\dot{J}_\parallel$, 
	where the torque $\dot{J}_\parallel$ is due to H$_2$ formation
	with a density of active sites $10^{13}\cm^{-2}\cm^{-2}$
(From LD99b).}

\end{figure}

The first striking question is why an effect so feeble as nuclear magnetism
can be so important in terms of internal relaxation. 
The answer 
is that spins align in a rotating body because of their
angular momentum, not because of their magnetic moments.
The magnetic moment enters only as a means for the spins to exchange
angular momentum with the lattice.
The coupling within the
electron spin system 
is very effective,
with the result that there is minimal ``lag'' in the electron spin
alignment when the grain angular velocity $\bOmega$ changes in grain
body coordinates.
In the case of the nuclear spin system, however, the value of the nuclear
magnetic moment is large enough to provide significant
 coupling, 
but weak enough so that there is a significant
lag\footnote{Purcell (1969) was the first to notice that for thermally
rotating $10^{-5}$~cm grains the rate of paramagnetic alignment in
the external magnetic field does not depend on the value of magnetic
moment of the paramagnetic species. Therefore he concluded that the
alignment of grains by interstellar magnetic field
may happen due to nuclear spins only.}
 in the nuclear spin alignment when $\bOmega$ precesses around the
grain axis of maximal inertia: 
the coupling is ``just right'' for nuclear relaxation
to be extremely effective for $10^{-5}\cm\la a \la 10^{-4}\cm$ 
grains.
The efficiency of nuclear relaxation drops for sufficiently high
frequencies and therefore
Barnett relaxation dominates for $a\la5\times10^{-6}\cm$,
as seen in Fig.~(4a).

As we discussed in \S3, LD99a
found that ``thermal flipping'' was an
important 
element
of grain dynamics for sufficiently small grains.
LD99b have shown that internal relaxation associated
with nuclear spin alignment
can be many orders of magnitude more rapid than due to the Barnett effect
for $a\ga 10^{-5}\cm$ grains.
As a result, the phenomena of thermal flipping and thermal trapping 
become important for grains as large as $\sim\!10^{-4}\cm$.

The Barnett effect is related to yet another quantum process 
relevant to the 
paramagnetic alignment of ultra-small grains ($<10^{-7}$~cm)
that is discussed in Lazarian \& Draine (2000).
Those grains produce important 
CMB foreground (see Fig. 4) which can be polarized if the grains
are aligned. 

Introduced by Davis \&
Greenstein (1951) paramagnetic relaxation is easy to understand: for a spinning
grain the component of interstellar magnetic field perpendicular to the grain angular velocity varies in grain coordinate. The resulting time-dependent
magnetization has associated energy dissipation and torques that damp grain
rotation perpendicular to magnetic field.

Rotation removes the spin degeneracy of the electron energy levels. 
The energy difference between 
electron spin parallel or antiparallel to $\vec \Omega$
provides a level splitting corresponding to $\hbar \omega = 
g\mu_{\rm B}H_{\rm BE}$. 

Now consider a (weak) static magnetic field $\bf H$ 
at an angle $\theta$ to $\vec \Omega$.
In grain coordinates, this appears like a static field $H\cos\theta$
plus a field $H\sin\theta$ rotating with frequency $\omega$.
This rotating field 
can be resonantly absorbed, since the energy level 
splitting is exactly $\hbar \omega$.

This energy absorption, of course, is proportional to 
paramagnetic susceptibility $\chi^{\prime\prime}(\omega)$.
In the classical Davis-Greenstein analysis this magnetic susceptibility
$\chi(\omega)$ is taken to be that of a sample at rest in zero magnetic
field: $\chi\equiv \chi(H_0=0, \omega)$. Here we point out that one should
instead use $\chi=\chi_{\bot}(H_0=H_{\rm BE}, \omega)$, where 
$\chi_{\bot}(H_0,\omega)$ describes the response of nonrotating material 
to a weak field rotating at
frequency $\omega$ perpendicular to a static magnetic field $H_0$.
Compared to classical Davis-Greenstein relaxation the study by
Lazarian \& Draine (2000) predicts that the relaxation happens at
the maximal possible or resonance rate. Therefore the relaxation
was termed ``resonance relaxation''. For grains of the order of
$10^{-5}$~cm this results in a correction factor of order unity.
For ultra-small grains it provides an enhancement of many orders of
magnitude.

\section{Is turbulence important for driving grains?}

Depending on grain relative velocities, grain-grain collisions can have various outcomes, e.g., coagulation, cratering, shattering, vaporization and erosion and ejection of mantle material (see Draine 1985 and references therein). It is likely that some features of grain size distribution (Mathis, Rumpl \& Nordsieck 1977; Kim, Martin \& Hendry 1994), eg., cutoff at large size, are the result of fragmentation.

Dust dynamics is also influential to the metallicity in various 
astrophysical environment. The depletion of heavy elements in interstellar 
medium is a long puzzling problem. Grains moving supersonically may 
efficiently accrete gas-phase heavy elements  (Weingartner \& Draine 1999, 
Wakker \& Mathis 2000). Dust ejection from galaxies is important 
to metal enrichment of intergalactic medium (see
Aguirre et al. 2001 and references therein). 

Various processes can affect the velocities of dust grains. Shocks, 
radiation, ambipolar diffusion, and gravitational sedimentation all can 
bring about a dispersion in grain velocities (Draine 1985).

{\it Shock acceleration}.~~~~~
The basic idea is that the weakly charged grains are like ions with
high mass to charge ratio (Epstein 1980). Thus they can easily diffuse farther back
upstream of the shock and be accelerated more efficiently to suprathermal
energies. Nevertheless, the shock acceleration is inefficient for low
speed grains. The reason is that the efficiency of the shock acceleration
depends on the scattering rate, which is determined by the stochastic
interaction with turbulence. For low speed particles, the scattering
rate is lower than the rate 
of momentum diffusion. In this case the stochastic acceleration
 by turbulence happens faster than dust acceleration 
by shocks (Park \& Petrosian 1996,
Yan \& Lazarian 2003, henceforth YL03, see also below). 

{\it Effects of radiation fields}.~~~~~
Grains are exposed to various forces in anisotropic radiation fields. 
Apart from
radiation pressure, grains are subjected to forces due to the asymmetric
photon-stimulated ejection of particles. A detailed discussion can
be found in Weingartner \& Draine (2001a, henceforth WD01; also the review 
by Draine 2003). The photoelectric
force depends on the ambient conditions relevant to
grain charging (WD01; Lafon 1990; Kerker \& Wang 1982). The calculation
by WD01 demonstrated that it is comparable to radiation pressure when
the grain potential is low and the radiation spectrum is hard. Photodesorption
is also a photon-stimulated ejection process, but of absorbed atoms
on grain surface. However, because the surface physics and chemistry
of grain materials are unclear, the calculation of the photondesorption
force is highly uncertain (WD01). The bottom line is that the force due to photodesorption
is expected to be comparable to the radiation pressure and photoelectric thrust  (Draine 2003).

{\it Motions arising from H$_{2}$ thrust}.~~~~~
A different residual imbalance arises from the difference of
the number of catalytic active sites for H\( _{2} \) formation on
opposite grain surfaces. The nascent H\( _{2} \) molecules leave
the active sites with kinetic energy \( E \), and the grain experiences
a push in the opposite direction. The uncompensated
force above is parallel to the rotational axis as the other components
of force are averaged out due to grain fast rotation. Applying
the characteristic values%
\footnote{The number of H$_{2}$ formation sites is highly uncertain. It may
also depend on the interplay of the processes of photodesorption and
poisoning (Lazarian 1995b; 1996). %
}  in Lazarian \& Draine (1997), LY02  
got the {}``optimistic'' velocity 
$v\simeq330(10^{-5}$cm$/a)^{1/2}$cm/s
for CNM (Fig.~5a) and $v\simeq370(10^{-5}$cm$/a)^{0.7}$cm/s for WNM.
The grains tend to be aligned with rotational axes parallel to the
magnetic field. Thus grains acquire velocities along the magnetic
field lines. It is clear from Fig.~5a that for the chosen set of parameters
the effect of H$_{2}$ thrust is limited. The percentage of atomic hydrogen is reduced in dark clouds, and the
radiation field is weak. Therefore, the velocities driven by the variation
of the accommodation coefficient are always much smaller than those
due to turbulence so that grains in dark clouds should be fully mixed.

The interstellar medium is turbulent (see Arons \& Max 1975; Scalo 1987; Lazarian 1999). Turbulence has been invoked by a number of authors (see Kusaka et al. 1970; V\(\ddot {o}\)lk et al. 1980; Draine 1985; Ossenkopf 1993; Weidenschilling \& Ruzmaikina 1994) to provide substantial grain relative motions. 

{\bf Grain Motions due to Frictional Drag}\\
In hydrodynamic turbulence, the grain motions are caused by the frictional
interaction with the gas. At large scale grains are coupled with the
ambient gas, and the slowing fluctuating gas motions mostly cause
an overall advection of the grains with the gas (Draine 1985). At
small scales grains are decoupled. The largest velocity difference
occurs on the largest scale where grains are still decoupled. Thus
the characteristic velocity of a grain in respect to the gas corresponds
to the velocity dispersion of the turbulence on the time scale $t_{drag}$
(Draine \& Salpeter 1979). Using the Kolmogorov scaling
relation $v_{k}\propto k^{-1/3}$, Draine (1985) obtained the largest
velocity dispersion in hydrodynamic turbulence $v\simeq V(t_{drag}/\tau_{max})^{1/3}$, where $\tau_{max}$ is the time scale of the turbulence at the injection scale.

As most interstellar are magnetized and magnetohydrodynamic (MHD) turbulence is much better understood now, a revisit on the problem, namely acceleration by MHD turbulence, is needed.
 In the following, we shall first introduce some basic idea of MHD 
turbulence and then give a brief review of our work on dust dynamics in MHD turbulence.
MHD perturbations can be decomposed into Alfv\'{e}nic, slow and fast
modes (see Alfv\'{e}n \& F\"{a}lthmmar 1963). Alfv\'{e}nic turbulence
is considered by many authors as the default model of interstellar
turbulence. Unlike hydrodynamic turbulence, Alfv\'{e}nic turbulence
is anisotropic, with eddies elongated along the magnetic field. This
happens because it is easier to mix the magnetic field lines perpendicular
to the direction of the magnetic field rather than to bend them. As
eddies mix the magnetic field lines at the rate $k_{\bot}v_{k}$ ,
where $v_{k}$ is the mixing velocity at this scale, the magnetic
perturbations (waves) propagate along the magnetic field lines at
the rate $k_{\parallel}V_{A}$ . The Alfv\'{e}n and slow modes can be described
by GS95 model (Goldreich \& Sridhar 1995; Cho \& Lazarian 2002; Lithwick \& Goldreich 2001; 
see also Cho, Lazarian \& Vishniac 2002 for a review). 
The corner stone of the GS95 model is a critical balance
between these rates, i.e., $k_{\bot}v_{k}\sim k_{\parallel}V_{A}$,
which may be also viewed as coupling of eddies perpendicular to the
magnetic field and wave-like motions parallel to the magnetic field.
Calculations by Cho \& Lazarian (2002; 2003c) demonstrated that fast modes are very similar to acoustic turbulence.

In the MHD case, grain motions are affected by magnetic fields. The charged
grains are subjected to the electromagnetic forces which depend on
the grain charge. If the periods of Larmor motion $\tau_{L}$
are longer than the gas drag time $t_{drag}$, the grains do not feel
magnetic field directly. Otherwise, grain perpendicular motions are constrained by magnetic
field.

As Alfvenic turbulence is anisotropic, it is convenient to consider
separately grain motions parallel and perpendicular to the magnetic
field. The corresponding discussions can be found in LY02, Lazarian
\& Yan 2002b.

The velocity dispersion induced by the compressional motion associated
with the fast modes also causes the relative movement of the grain
to the ambient gas (see YL03). 
The velocity fluctuations associated with
fast modes are always in the direction perpendicular to $\mathbf{B}$
in a low $\beta$ ($\beta\equiv P_{gas}/P_{mag}$) medium. Thus the grain velocities are also perpendicular
to $\mathbf{B}$. In high $\beta$ medium, grains can have velocity
dispersion in any direction as the velocity dispersions of fast modes
are radial, i.e., along $\mathbf{k}$.

{\bf Acceleration of Grains by Gyroresonance}\\
In YL03 we identified 
a new mechanism of grain acceleration, namely, gyroresonance
 that is based on the direct interaction of charged grains with MHD turbulence. 
There exists an important
analogy between dynamics of charged grains and dynamics of cosmic rays
(see Yan \& Lazarian 2002), and the existing machinery developed for cosmic rays was
 modified to describe charged grain dynamics. The energy exchange
involves resonant interactions between the grains and the waves (see
YL03). Specifically, the resonance condition is $\omega-k_{\parallel}v\mu=n\Omega$,
($n=0,\pm1,\pm2...$), which means that the Doppler-shifted frequency of the wave 
in the grain's guiding center
rest frame $\omega_{gc}=\omega-k_{\parallel}v\mu$ is a multiple of
the particle gyrofrequency $\Omega$. Basically there are two main types
of resonant interactions: gyroresonance acceleration and transit acceleration.
Transit acceleration ($n=0$) requires longitudinal motions and only
operates with compressible modes. As the dispersion relation is $\omega=kV_{f}>kV_{A}$
for fast waves, it is clear that it can be only applicable to super-Alfvenic
(for a low $\beta$ medium) or supersonic (for a high $\beta$ medium)
grains. For low speed grains that we deal with here, gyroresonance is the dominant MHD interaction.

The calculation by YL03 showed grains gain the maximum velocities perpendicular
to the magnetic field and therefore the averaged $\mu$ decreases.
This is understandable since the electric field which accelerates
the grain is in the direction perpendicular to the magnetic field.

Then we applied our results to various idealized phases of interstellar
medium (YLD03). In Fig.~5a, we show the velocity of grain 
as a function of grain size in CNM. 

\begin{figure}
{\centering \leavevmode
\epsfxsize=2.0in\epsfbox{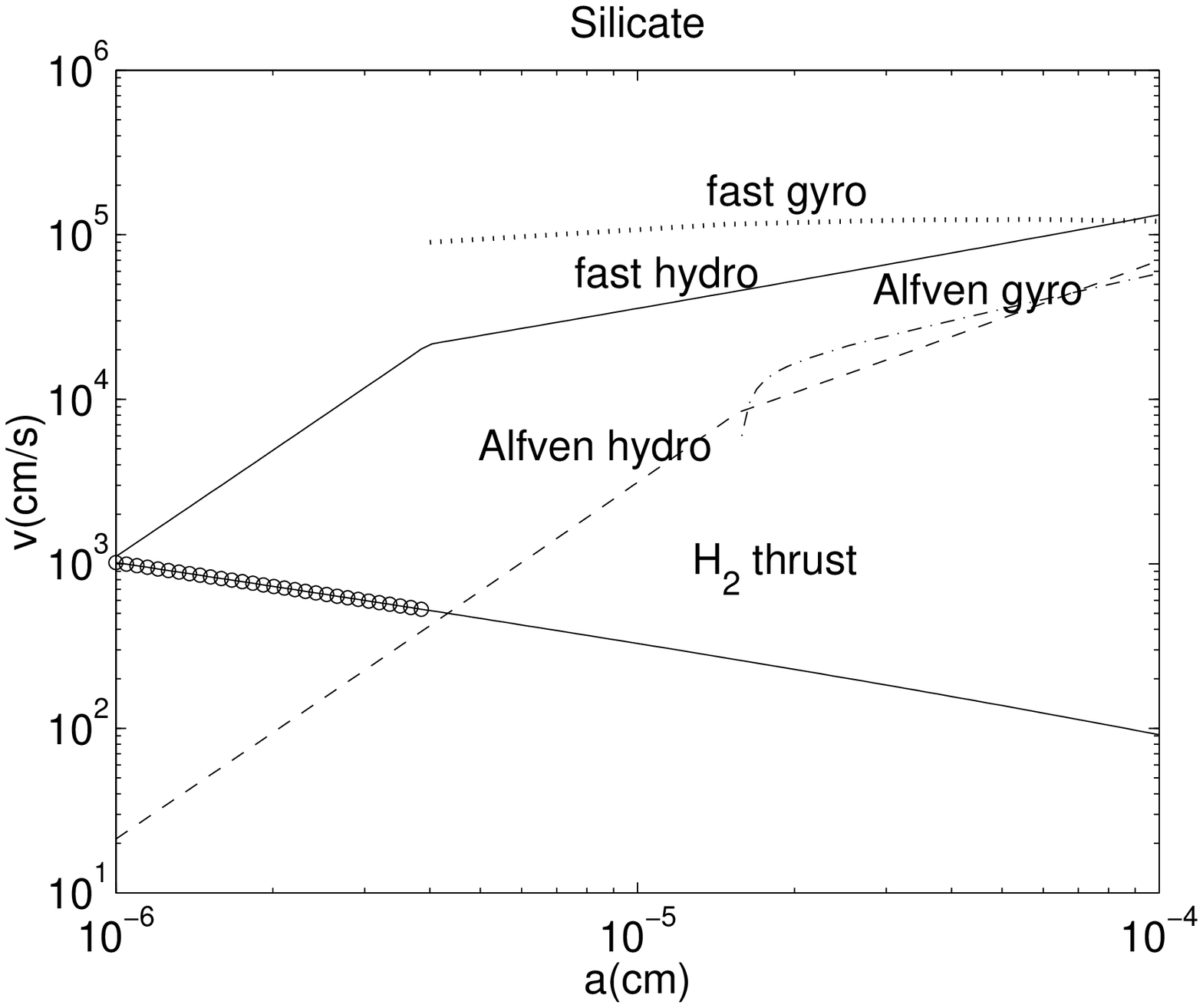}
\hfil
\epsfxsize=2.0in\epsfbox{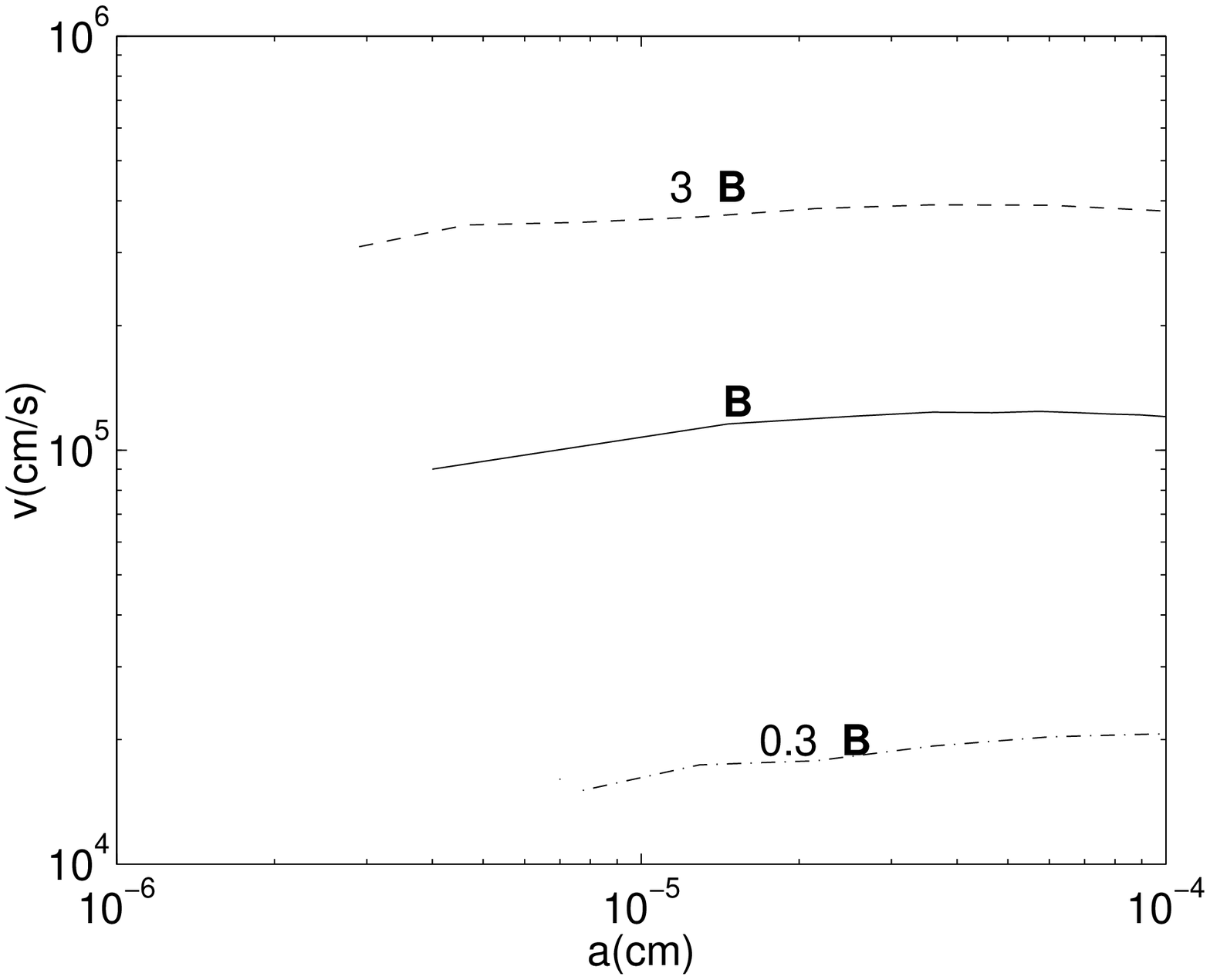}}

\caption{{\it Left panel}-- 
Relative velocities as a function of radii in CNM
for silicate grains. The dotted
line represents the gyroresonance with fast modes. The dashdot line
refers to the gyroresonance with Alfv\'{e}n modes. The cutoff is by
viscous damping. The dashed line is the result from hydro drag with
Alfv\'{e}n modes (see LY02), the solid line represents the hydro
drag with fast modes. Contributions from different mechanisms are
approximately additive in squares.
The grain velocity driven by H$_{2}$ formation (solid line). The
part marked by open circles is nonphysical because thermal flipping
is not taken into account (YL03). {\it Right panel}-- Grain velocities
in CNM
 gained from gyroresonance for different magnetic field 
strengths (YLD03).}
\end{figure}

The acceleration by gyroresonance in both MC and DC are not so efficient
as in other media. This happens because the time for the gyroresonant
acceleration $t_{drag}$ are much shorter in MC and DC. In these media, the
drag time because of the high density is less than the gyro-period
for grains larger than $10^{-5}$cm.

It should be noted that the strength of magnetic fields in ISM is
still somewhat uncertain and may vary from place to place. We adopted
a particular set of values in above calculations. How would the results
vary as the magnetic field strength vary? First of all, we know that
the critical condition for acceleration is that the Larmor period
$\tau_{L}$ is greater than the period of the turbulent motions at the
damping scale. Grains 
with $\tau_{L}$ less than that can not be accelerated via gyroresonance. 
Thus the cutoff
grain mass $m_{c}$ varies with the medium, $m_{c}\sim q\tau_{c}B/c$. The magnitude of the velocity is a complex function of
the magnetic field. For illustration, here we demonstrate the results
for 3 times stronger and weaker magnetic field (see Fig.~5b). Since 
the acceleration via hydro drag by fast modes decreases with the magnetic field, the relative importance of
 the two processes, namely, gyroresonance
and hydro drag depends on the magnitude of the magnetic field. In
magnetically dominant region, gyroresonance is dominant. In weakly
magnetized region, the decoupled motions provide the highest acceleration
rate. The injection scale is another uncertain parameter, but the
grain velocity is not so sensitive to it provided that the injection
scale is much larger than the damping scale.

\section{Discussion}

{\bf Implications of Rotational Dynamics}

\emph{Grain Alignment}

Grain alignment critically depends on rotational dynamics of grains.
Incorporation of thermal fluctuations within theories of grain alignment
(Lazarian 1994, 1997ab, Lazarian \& Roberge 1997, Lazarian \& Draine 1997, 
Roberge \& Lazarian 1999, Weingartner \& Draine 2003) 
introduced substantial changes in our understanding
how grains get aligned. Thermal flips (LD99a,b) and thermal trapping 
(LD99a) that they entail have radically changed some of the contemporary 
paradigms of grain alignment. 
Indeed, if grains get thermally trapped as discussed earlier, 
collisions with gaseous atoms randomize vectors of angular
momenta to high degree. Due to efficiency of nuclear relaxation
(LD99b) most of the grains do not rotate suprathermally in the absence
of radiative torques. This explains why small grains for which radiative
torques are not important are only marginally aligned. An alternative
explanation of this fact is given by Mathis (1986) and is related to
the preferential presence of ferromagnetic inclusions within large
grains. However, it is claimed in Lazarian (2003) that data in
Whittet et al. (2001) is consistent only with the change of the alignment
efficiency with the extinction rather than the grain size. More
discussion of grain alignment can be found in the paper by Roberge (2004)
in this volume.

\emph{Structure of Grains}

While grain formation may favor loosely bound conglomerates of low fractal
dimension, rapid rotation of grains destroys grain that are not 
sufficiently compact. The tensile stress calculated in Draine \& Lazarian (1998b)
did not place severe constrains on small grain considered unless the 
grains consist of loosely connected parts.

{\bf Implications of Translational Motion} 

\emph{Shattering and Coagulation}

With the grain relative velocities known, we can make predictions
for grain shattering and coagulation. For shattering, we adopt the
Jones' et al. (1996) results, namely, the shattering threshold is
2.7km/s for silicate grain and 1.2km/s for carbonaceous grain. The
critical sticking velocity were calculated in Chokshi et al. (1993)
(see also Dominik \& Tielens 1997).%
\footnote{There are apparent misprints in the numerical coefficient of Eq.(7)
in Chokshi et al.(1993) and the power index of Young's modulus in
Eq.(28) of Dominik \& Tielens (1997). %
} However, experimental work by Blum (2000) shows that the critical
velocity is an order of magnitude larger than the theoretical calculation.
Comparing these critical velocities with the velocity curve we obtained
for various media, we can get the corresponding critical size for
each of them (see Table 1). However, given the uncertainties
with the parameters as discussed above, we are cautious about these
numbers.

\begin{table}

\begin{tabular}{|c|cc|c|c|c|c|}
\hline 
ISM&
 CNM&
 CNM&
  WNM&
 WNM&
 WIM&
 WIM\tabularnewline
\hline
Material&
  Si&
  C&
  Si&
 C&
 Si&
 C\tabularnewline
\hline
Shattering size ($\mu$m)&
	NA&
	NA&
  $>0.2$&
 $>0.2$&
$>0.003$&
 $>0.001$\tabularnewline
\hline
Coagulation size ($\mu$m)&
 $<0.01$&
   $<0.02$&
 $<0.02$&
 $<0.05$&
NA&
NA \tabularnewline
\hline
\end{tabular}
\caption{The critical shattering and coagulation size in different medium. NA=not applicable.}
\end{table}

\emph{Correlation between turbulence and grain sizes}

Change in the intensity of turbulence should entail variations in
grain sizes. The grain velocities are strongly dependent on the maximal
velocity of turbulence $V$ at the injection scale, which is highly
uncertain. Thus the critical coagulation and shattering sizes would
also depend on the amplitude of the turbulence accordingly.

\emph{Composition of cosmic ray}

It has been shown that the composition of the galactic cosmic ray
seems to be better correlated with volatility of elements (Ellison,
Drury \& Meyer 1997). The more refractory elements are systematically
overabundant relative to the more volatile ones. This suggests that
the material locked in grains must be accelerated more efficiently
than gas-phase ions. This leads to
the speculation of acceleration of grain erosion products in shocks
(Epstein 1980; Cesarsky \& Bibring 1981; Bibring \& Cesarsky 1981).
 If the sputtering happens upstreams, the sputtered products will
be carried downstream, where they can be further accelerated to
cosmic ray energies with higher efficiencies than gas-phase ions (Epstein
1980). The destruction by shocks may also be a prodigious source of
PAHs, HACs and small grains (Jones, Tielens \& Hollenbach 1996, Tielens
et al. 1994). The stochastic acceleration is more efficient for low speed grains
and thus can serve as a preacceleration mechanism for shock acceleration. 

\emph{Vacuum cleaning and mechanical alignment}

Our results indicate that grains can get supersonic through interaction
with fast modes. Grains moving supersonically can efficiently vacuum-clean
heavy elements as suggested by observations (Wakker \& Mathis 2000).
The supersonic grains can also be aligned (see a review by Lazarian
2003 and references therein). As pointed out earlier, the scattering
is not efficient for slowly moving grains so that we may ignore the
effect of scattering to the angular distribution of the grains. Since
the acceleration of grains increases with the pitch angle of the grain, 
the supersonic grain motions
will result in grain alignment with long axes perpendicular to the
magnetic field.

\emph{Grain Segregation and Turbulent Mixing}

Our results are also relevant to grain segregation. Grains are the
major carrier of heavy elements in the ISM. The issue of grain segregation
may have significant influence on the ISM metallicity. Subjected to
external forcing, e.g., due to radiation pressure\footnote{Even in the
absence of radiation pressure grains can move along magnetic field
lines due to the uncompensated forces, e.g. due to an unequal
number of active sites of H$_2$ formation (see P79). Those forces would
be mitigated in molecular clouds, which would induce inflow of dust
into molecular cloud. The latter would affect metallicity of the 
newborn stars.}, grains gain size-dependent
velocities with respect to gas. WD01 have considered the forces on
dust grains exposed to anisotropic interstellar radiation fields.
They included photoelectric emission, photodesorption as well as radiation
pressure, and calculated the drift velocity for grains of different
sizes. The velocities they got for silicate grains in the CNM range
from $0.1$cm/s to $10^{3}$cm/s. Fig.~5a shows that the turbulence
produces larger velocity dispersions\footnote{Our calculation show that for
the chosen set of parameters the effects of H$_{2}$ thrust are also
limited.}. Those velocities are preferentially perpendicular to magnetic
field, but in many cases the dispersion of velocities parallel\footnote{This
dispersion stems from both the fact that the transpositions of matter
by fast modes are not exactly perpendicular to magnetic field (see plot
in Lazarian \& Yan 2002b) and due to randomization of directions of
grain velocities by magnetized turbulence (Yan \& Lazarian 2003).} to
magnetic field will be comparable or greater than the regular velocities above.

More important is that
if reconnection is fast (see Lazarian \& Vishniac 1999), the mixing
of grains over large scales is provided by turbulent diffusivity $\sim VL$.
Usually it was assumed that the magnetic fields strongly suppress
the diffusion of charged species perpendicular to their directions. However,
this assumption is questionable if we notice that motions perpendicular
to the local magnetic field are hydrodynamic to high order as suggested
by Cho, Lazarian \& Vishniac (2002). In fact, recent work by Cho et
al. (2003) shows that the diffusion processes in MHD turbulence are
as efficient as in hydrodynamic case if the mean magnetic field is
weak or moderately strong, i.e., $\mathbf{B}_{0}$ is $\sim$ equipartition
value. This means that
 from the theoretical perspective, we do expect that grains
can be mixed by the MHD turbulence. Therefore we believe that
the segregation of very small and large grains
speculated in de Oliveira-Costa et al. (2002) is unlikely to happen
for typical interstellar conditions.

{\bf Interdependence of Rotational and Translational Motions}

Coupling of different types of motions is very important for
understanding grain dynamics. As we discussed earlier, the coupling
of vibrational and rotational motions results in flips that change
dynamics of grains.

Similarly translational and rotational motions are interdependent.
For instance, frequent flips average out not only uncompensated
torques, but also the uncompensated force. 
As we mentioned earlier, radiative torques can
rescue grains from thermal trapping and therefore we show in Fig~5
that grains larger than $\sim 5\times 10^{-6}$~cm experience 
uncompensated forces due to H$_2$ formation.

Fast rotation makes atoms less susceptible to mechanical alignment through
gaseous collisions. However, as grains undergo crossovers, they can still
be aligned via processes that were termed in Lazarian (1995a) crossover
and cross sectional alignment (see also Lazarian \& Efroimsky 1996; 
Efroimsky 2002).  

Streaming grains get rotation rates that are faster than 
that of their thermally rotating counterparts.
This may allow thermally trapped grain get ``untrapped'' and
therefore acquire suprathermal rotation through uncompensated
Purcell's torques (P79).
If grains are ``helical'' in terms of collisions with gaseous atoms the effect
of grain spinning up and getting untrapped should be even move pronounced.

In addition, some characteristics of dust grains are affected both
by rotational and translational motions. For instance, both fast
rotation and collisions limit the structure of grains and prevent very loose
aggregates to dominate the extinction.

\section{Summary}

~~~1. Interstellar grains move relatively to each other and this results
in collisions. The outcome of those may be shattering or coagulation
depending on the relative velocities of grains. As grains move relatively
to gas they adsorb atoms from the gas and this affects dust chemistry.
In addition grains get aligned as a result of gas- dust streaming.
Grain rotation affects both grain alignment and grain size distribution.
Fast rotating grains are not subjected to randomization via gas
collisions. Fast rotation destroys loosely connected
grains.

2. Grains do not always rotate about their axis of major inertia.
Thermal fluctuations cause vibrations of grain angular momentum
about the axis of major inertia and occasional flips. The rate
of flips depends on the coupling of rotational and vibrational
degrees of freedom that is determined by the rate of dissipative
processes in the wobbling grain. Frequent flips result in grains
being {\it thermally trapped}, i.e. rotating with thermal velocities
in spite of the presence of suprathermal torques.

3. The rate of grain rotation is determined by collisions with
neutrals, ions, photons, plasma effects, emission of radiation,
nascent H$_2$ molecules etc. It is also influenced by the
grain flipping. Depending on the interplay of these processes,
grains can rotate both at a rate that is larger and smaller
than the rate of the thermal Brownian motion. 

4. Although rotation is classical even for the smallest grains, a number
of subtle quantum mechanical
effects determine grain dynamics. For instance, plasma
drag arising from the interaction of the grain dipole moment with
impinging ions is limited by the quantum nature of the interactions.
Nuclear internal relaxation is a quantum effect arising from reorienting
of spins of nuclei within a wobbling grain. Paradoxically, in terms
of coupling of rotational and vibrational degrees of freedom nuclear
spins can be much more efficient than their electron counterparts.
Resonance paramagnetic relaxation is yet another quantum effect that
alleviates alignment of tiny grains. 

5. Grains can be accelerated by various mechanisms. Formation of
H$_2$, variations of the accommodation coefficient, photoelectric emission 
and photodesorption,
radiation pressure are the processes that can induce grain motion.
However, ubiquitous interstellar turbulence is probably the major
driver of relative grain-gas and grain-grain motions. As the turbulence
is usually magnetized and grains are usually charged the magnetic 
nature of turbulence is essential. Magnetic fields, on one hand, limit
the relative gas-grain motions in the direction perpendicular to 
the local magnetic field. On the other hand, interaction of charged
grains with magnetic turbulence results in resonance acceleration of
dust. Supersonic grain velocities are attainable as a result
of such an interaction. 

{\bf Acknowledgment} The NSF grants NSF AST-0098597 NSF AST -0243156 are
acknowledged. The work on grain dynamics in magnetized turbulence is
partially supported by the NSF Center for magnetic self-organization in
astrophysical and laboratory plasmas at the University of Wisconsin-Madison.
We thank John Mathis and Bruce Draine 
for reading the manuscript and valuable comments.

\end{document}